\documentclass[journal]{IEEEtran}
\IEEEoverridecommandlockouts
\usepackage{lineno}
\usepackage{subfigure}

\usepackage{graphicx}
\usepackage{amsmath}
\usepackage{cite}
\usepackage{float}
\usepackage{lettrine}
\usepackage{soul}
\usepackage{color}
\usepackage[table]{xcolor}
\usepackage{url}
\usepackage{diagbox}
\usepackage{multirow}
\usepackage{array}
\usepackage[ruled]{algorithm2e}
\usepackage{enumitem}
\usepackage{xcolor}
\usepackage{xspace}
\sethlcolor{yellow}

\newcolumntype{M}[1]{>{\centering\arraybackslash}m{#1}}
\newcolumntype{N}[1]{>{\raggedright\arraybackslash}m{#1}}
\newcommand{\eat}[1]

\usepackage{tikz}

\usepackage{booktabs}

\def\SDVSC{SDViSC\xspace}
\def\SDVSCs{SDViSCs\xspace}
\def\VSC{ViSC\xspace}
\def\VSCs{ViSCs\xspace}

\graphicspath{ {fig/} }

\begin{document}

\title{Software-Defined Virtual Synchronous Condenser}

\author{Zimin~Jiang,
        Peng~Zhang,
        Yifan~Zhou, Łukasz~Kocewiak,
        Divya~Kurthakoti~Chandrashekhara,
        Marie-Lou~Picherit,
        Zefan~Tang,   Kenneth B. Bowes,
        and Guangya Yang
\thanks{
}

}

\maketitle

\begin{abstract}
Synchronous condensers (SCs) play important roles in integrating wind energy into relatively weak power grids. 
However, the design of SCs usually depends on specific application requirements and may not be adaptive enough to the frequently-changing grid conditions caused by the transition from conventional to renewable power generation. This paper devises a software-defined virtual synchronous condenser (\SDVSC) method to address the challenges. Our contributions are fourfold: 1)  design of a virtual synchronous condenser (\VSC) to enable full converter wind turbines to provide built-in SC functionalities; 2) engineering \SDVSCs to transfer hardware-based \VSC controllers into software services, where a Tustin transformation-based software-defined control algorithm guarantees accurate tracking of fast dynamics under limited communication bandwidth; 3) a software-defined networking-enhanced \SDVSC communication scheme to allow enhanced communication reliability and reduced communication bandwidth occupation; and 4) Prototype of \SDVSC on our real-time, cyber-in-the-loop digital twin of large-wind-farm in an RTDS environment. Extensive test results validate the excellent performance of \SDVSC to support reliable and resilient operations of wind farms under various physical and cyber conditions. 


\end{abstract}


\section{Introduction}

\IEEEPARstart{O}{ffshore} wind energy has been increasingly integrated into power grids. For instance, the State of New York has set the goal of integrating 9 GW offshore wind by 2035 to help build 100\% electricity grids powered by zero-emission resources~\cite{sunrise,liu2022open,wan2022cooperative}. Large scale integration of offshore wind energy poses challenges to the operation of weakened power grid~\cite{wu2019impedance,bakhshizadeh2018improving,li2022dominant}.


One effective approach to support large-scale integration of offshore wind energy is to use synchronous condensers (SCs), which can provide inertia~\cite{nguyen2018combination}, frequency regulation support~\cite{nguyen2020applying}, short-circuit current contribution~\cite{jia2018impact}, and sub/super-synchronous oscillations suppression~\cite{wang2020impact} to the grid. 
However, conventional SCs are synchronous machines and their designs (e.g., locations and capacities) highly depend on the application scenarios~\cite{nguyen2021technical,bao2022wind}. 
Correspondingly, the deployment and upgrade of SCs are highly expensive and they are unable to 
adapt to the frequently-changing grid conditions~\cite{jia2018impact}. 




The overarching goal of this paper is to originate a lightweight, cost-effective, and adaptive approach to achieve the SC functionalities for wind energy integration. On the one hand, at the wind turbine and wind farm level, inverter controllers can provide built-in capabilities to realize various control functionalities, such as hierarchical control, synthetic inertia and instability damping~\cite{kocewiak2020overview}. 
Intuitively, if the inverter controllers are properly designed, they will be able to provide the desired SC functionalities while avoiding deploying the real synchronous machine hardware.  
On the other hand, software-defined control (SDC)~\cite{wang2020software}, a recently emerging technology, also provides insights into resolving the hardware dependence issue. SDC virtualizes traditionally hardware-dependent inverter controllers into software-defined services and, in this way, realizes ultra-flexible and cost-effective controllers for power grids with significantly enhanced programmability and deployability. 

In summary, this paper develops a software-defined virtual synchronous condenser (\SDVSC), which realizes the SC functions without requiring the deployment of real SCs and hardware-dependent controllers.  
The main contributions are: 
\vspace{-2pt}
\begin{itemize}[leftmargin=*]
    \item \textbf{Architecture of virtual synchronous condenser (\VSC)}: We enable a built-in capability of wind turbines to provide programmable SC functions such as flexible Var capability,  inertia, oscillation damping, and weak grid enhancement. 
    \item \textbf{Engineering \SDVSC}: We accomplish \SDVSC by devising a Tustin transformation-based SDC algorithm to virtualize the hardware \VSC controller as a software service, which enables capturing the fast dynamics of \VSC under limited communication bandwidth. 
    \item \textbf{Resilient \SDVSC communication by SDN}: 
    We further establish software-defined networking (SDN)-enabled communication for \SDVSC to enable  reliable operations of \SDVSCs with low delays and high robustness even under communication network impairments.
    \item \textbf{Validated Prototype of \SDVSC}: A cyber-in-the-loop large-wind-farm prototype with 50 wind turbines is built in an RTDS environment, and extensive test results validate the superior programmability and flexibility of \SDVSC  
    and its capability to enable reliable operations of wind farms under weak grid conditions and cyber failures. 
\end{itemize}
\vspace{-2pt}

The remainder of this paper is organized as follows. Section II presents the design of \VSC and \SDVSC. Section III develops the SDN-enhanced \SDVSC. Section IV presents the established digital twin of \SDVSC and provides extensive  experiments to validate the performance of \SDVSC. Section V concludes the paper.

\vspace{-10pt}
\section{\SDVSC for Wind Energy Integration}

\begin{figure}[!h]
  \centering
\includegraphics[width=0.4\textwidth]{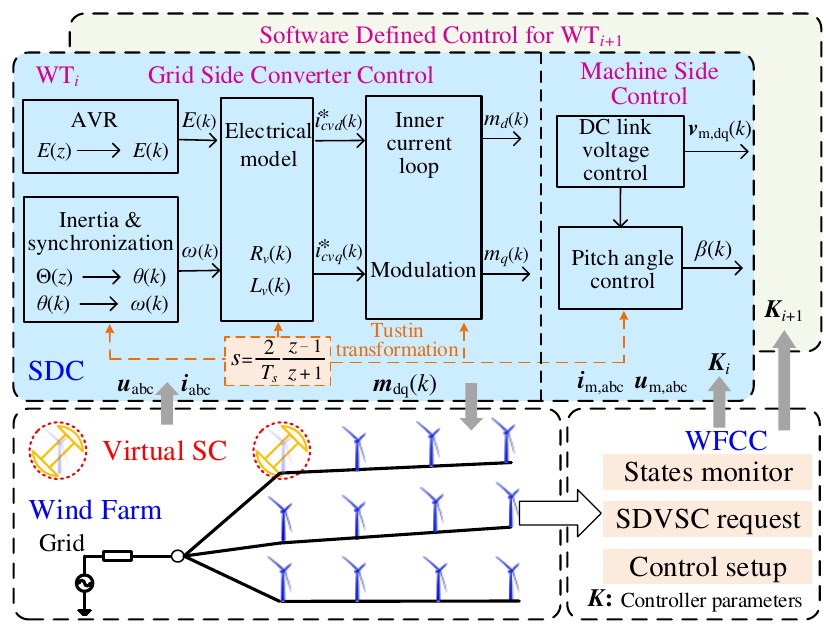}
  \vspace{-10pt}
\caption{\label{fig:SDC_Implement} Design overview of \SDVSC.}
  \vspace{-12pt}
\end{figure}

This section devises \SDVSC, a software-defined virtual synchronous condenser, as seen Fig.~\ref{fig:SDC_Implement}, to support wind energy integration in an unprecedentedly programmable, adaptative and lightweight manner. In this paper, ViSC refers to SC functionalities provided by inverters with controllers implemented on specific hardware, such as DSP or PLC; SDViSC means that SC functionalities are provided by inverters with pure software-based controllers.

\vspace{-10pt}
\vspace{-6pt}

\subsection{Architecture of \VSC}

We first design \VSC, which leverages wind generator converters to achieve the following control functions of SCs: i) to provide a natural inertial response and a short-circuit current, and ii) to generate or absorb reactive power to support voltage regulation. The design of \VSC includes control strategies for both grid-side and machine-side converters.


\subsubsection{Grid-Side Control}

\begin{figure}[!h]
  \centering
  \includegraphics[width=0.495\textwidth]{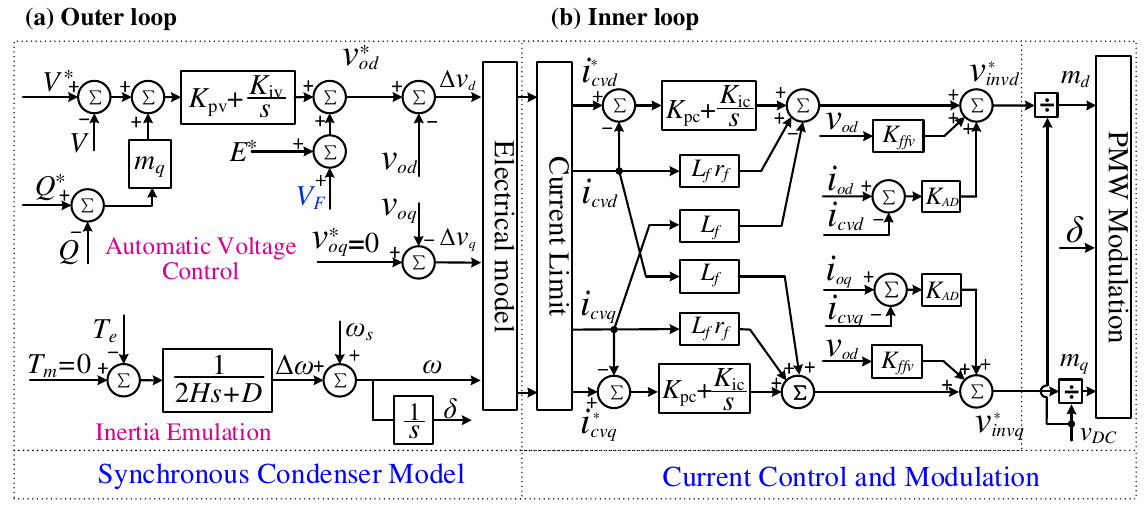}
  \caption{\label{fig:Fig2} Grid-side control of \VSC.}
  \vspace{-8pt}
\end{figure}

Our design of the grid-side converter adopts a double-loop control, as shown in Fig.~\ref{fig:Fig2}. 
\vspace{-8pt}
The outer-loop control is designed to achieve the following functionalities (see Fig.~\ref{fig:Fig2}(a)): 
i) a mechanical model to provide inertia emulation and synchronization with the grid; 
ii) a voltage controller to provide the automatic voltage regulation (AVR) as conventional SCs, where an electrical model represents the stator windings;
and iii) a supplementary controller customized to provide auxiliary control functions.




For the inertia emulation part, the transfer function to generate the phase angle $\delta$ is formulated as:
\vspace{-5pt}
\begin{equation}
     \Theta(s)  = \frac{1}{2Hs^2+Ds}T(s)+\frac{1}{s}W(s)
     \label{Eq:rotor_trans}
\end{equation}
where $\Theta(s)$, $T(s)$ and $W(s)$ are the Laplace transforms of the phase angle $\delta$, injected torque error $T_{err}$=$T_m$-$T_e$ ($T_m=0$ due to no input mechanical power), and the reference angular frequency $\omega_s$, respectively; $H$ and $D$ are the emulated inertia constant and the damping factor of the \VSC, respectively.

For the AVR part, its transfer function is formulated as:
\vspace{-5pt}
\begin{equation}
    E = [(V^*-V)+m_q(Q^*-Q)](K_{pv}+\frac{K_{iv}}{s})+E^*+V_F
    \label{Eq:AVR}
\end{equation}
where $K_{pv}$ and $K_{iv}$ are the PI parameters of AVR; and $m_q$ is the droop coefficient; $E$, $V$, $Q$, $E^*$, $V^*$ and $Q^*$ are the AVR output, output voltage, output reactive power and the corresponding references, respectively; $V_F$ is the output of the supplementary controller, which will be discussed in \eqref{eq:supplementary}.


The internally-induced voltage of the AVR block (i.e., $E$) is further connected with an electrical module to represent the electrical behavior of a synchronous machine. A commonly-adopted quasi-stationary electrical model is applied~\cite{mo2017evaluation}: 
\vspace{-5pt}
\begin{equation}
    \tilde{i}_{cvd}+j \tilde{i}_{cvq}=\frac{E-(v_d+jv_q)}{R_v+jX_v}
\end{equation}
where $R_v = R_{vir}/[(\omega L_{vir}/\omega_s)^2+R_{vir}^2]$ and $X_v = (\omega L_{vir}/\omega_s)/[(\omega L_{vir}/\omega_s)^2+R_{vir}^2]$ represent the dynamic impedance; $R_{vir}$ and $L_{vir}$ are the virtual resistance and inductance; $\omega$ is the angular frequency of \VSC. 
A circular current limiter~\cite{taul2019current} is applied to limit the amplitude and preserve the angle of the output of the electrical model $\Tilde{i}_{cvd}+j\Tilde{i}_{cvq}$.
The output of the current limiter, i.e., $i_{cvd}^*+ji_{cvq}^*$, provides the current reference for the inner-loop controller.

For the supplementary control block, in this paper, we design the following virtual friction control function to help \VSC suppress power oscillations and improve the frequency stability under weak grid conditions:  
\vspace{-5pt}
\begin{equation}\label{eq:supplementary}
    V_F(s) = \frac{W(s)-\Omega(s)}{1+sT_{LPF}}(\frac{K_{F1}s}{1+sT_{v1}}+\frac{K_{F2}s}{1+sT_{v2}})\Big (\frac{1+sT_1}{1+sT_2} \Big)^2
\end{equation}
where  $T_{LPF}$, $T_{V1}$, $T_{V2}$, $T_1$ $T_2$ and $K_{F1}$, $K_{F2}$ are the controller's time constants and gains; the input signal is the frequency deviation $(\omega_s -\omega)$, $\Omega(s)$ is the Laplace transform of $\omega$; the output signal $V_{F}$ is then sent to the AVR block as shown in (\ref{Eq:AVR}).
Eq. (\ref{eq:supplementary}) adopts two control signals, i.e., the frequency deviation and the rate of change of frequency, to damp the power oscillation. A lead-lag controller to adjust the output magnitude is also used to further help oscillation suppression due to its better trade-off between static accuracy and stability.

The inner loop of the grid-side converter, as detailed in Fig.~\ref{fig:Fig2}(b), includes a decoupled PI controller, a voltage feed-forward controller and an additional active damping controller, which improves the output current quality and suppresses filter voltage oscillations. 

\subsubsection{Machine-Side Control} 
\begin{figure}[!t]
  \centering
  \includegraphics[width=0.45\textwidth]{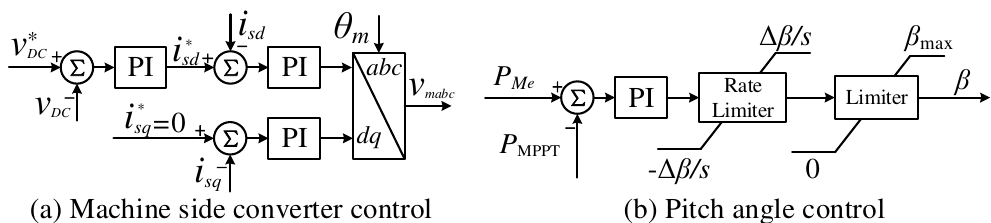}
  \vspace{-6pt}
    \caption{\label{fig:MSC_Control} Machine-side control.} 
  \vspace{-12pt}
\end{figure}

As for the machine-side control, our main target is to maintain a constant DC link voltage regardless of variations in the wind turbine machine speed $\omega_r$ and torque. This is achieved by the machine-side converter control and the pitch angle control, as shown in Fig.~\ref{fig:MSC_Control}.

For the machine-side converter control (see Fig.~\ref{fig:MSC_Control}(a)), the DC link voltage difference $(v_{DC}^*-v_{DC})$ is regulated by a PI controller to generate the stator $d$-axis current reference $i_{sd}^*$. The stator $q$-axis current reference $i_{sq}^*$ is set as 0, because the active power $P_{Me}$ generated by the machine can be expressed as $P_{Me} = \frac{2}{3}\psi \omega_r i_{sq}$ \cite{chinchilla2006control}, where $\psi$ is the stator magnetic flux linkage. 
The pitch angle control (see Fig.~\ref{fig:MSC_Control}(b)) is used to match the power $P_{Me}$ consumed by the machine to maintain constant DC link voltage. 
The difference between $P_{Me}$ and the captured wind power $P_{MPPT}$ using maximum power point tracking (MPPT) is sent to a PI controller and the reference pitch angle is obtained through the pitch angle control framework~\cite{lyu2019coordinated}.

Consequently, through 
\VSC, wind turbines can provide inertia support and voltage regulation as conventional SCs. Moreover, it provides programmable control performance to some extent, because the inertia constant and damping factor in corresponding control blocks can be flexibly adjusted.

\vspace{-4pt}
\subsection{\SDVSC: Software-Defined \VSC} \label{sec:SDVSC}
Further, we discuss how to engineer an \SDVSC, i.e., realizing \VSC 
in edge computing  to fully release the virtualization and programmability advantages. The key to the development is a Tustin transformation-based SDC approach.

The basic philosophy of SDC is to achieve hardware-dependent control functions in a software manner by discretizing all continuous-time controllers into discrete-time ones~\cite{wang2020software}. During the  SDC discretization procedure, conventional numerical integrators, such as Euler and trapezoidal methods, usually require small sampling rates to track fast dynamics to perform real-time control~\cite{zhao2021novel,wang2020software}, which unavoidably increases communication burdens from the SDC unit to WTs (see Section \ref{sec:SDNCom}), e.g., large communication bandwidth consumption. 








To bridge the gap, this subsection devises a Tustin transformation-based SDC algorithm to realize \SDVSC under fast dynamics and limited communication bandwidth.  

\subsubsection{Tustin Transformation}
The Tustin transformation is essentially a Taylor expansion-based ``$s$ to $z$" mapping of $z$=$e^{sT_s}$, as shown below:
\vspace{-5pt}
\begin{equation}
    s = \frac{1}{T_s}Ln(z) = \frac{2}{T_s}\sum_{k=1}^{\infty}\frac{1}{2k-1}(\frac{z-1}{z+1})^{2k-1}
\end{equation}
where $T_s$ is the sampling period. Using only the first term, the Tustin transformation is defined as follows:
\vspace{-5pt}
\begin{equation}
    s=\frac{2}{T_s}\frac{z-1}{z+1}
    \label{Eq:Tustin}
\end{equation}
The Tustin transformation preserves the properties of stability and the minimum phase of the original continuous-time controllers~\cite{al2007novel}. Thus, the original controllers' performance can be guaranteed after discretization without requiring tiny sampling rates as required by conventional integration rules~\cite{wang2020software}.

Next, we discretize \VSC using the Tustin transformation.

\subsubsection{Discretization of the AVR Block} \label{sec:Discretization}
PI controllers are the main-body of the AVR block. Given an arbitrary PI controller with the input $e_r(t)$, output $g(t)$ and parameters $K_p$ and $K_i$, the Tustin transformation can be performed as:
\vspace{-5pt}
\begin{equation}
    G(z)(1-z^{-1})=[\frac{T_s}{2}K_i(1+z^{-1})+K_p(1-z^{-1})]E_r(z)
\end{equation}
where $G(z)$ and $E_r(z)$ are the $z$ transforms of $g(t)$ and $e_r(t)$. The PI controller output can be calculated recursively as:
\vspace{-5pt}
\begin{align}
    g(k) & = g(k-1)+\frac{T_s}{2}K_i(e_r(k)+e_r(k-1)) +K_p(e_r(k)\notag \\ &-e_r(k-1))
    \label{PI_Controller}
\end{align}
\vspace{-16pt}

Consequently, the output voltage $E$ of the AVR controller is calculated by the following rule:
\vspace{-5pt}
\begin{align}
      E(k) & = E(k-1) + \frac{T_s}{2}K_{iv}[2(V^*+m_qQ^*)-V(k-1)\notag \\ &-V(k)-m_q(Q(k)+Q(k-1))]- K_{pv}(V(k) \notag \\ &-V(k+1)-m_q(Q(k)-Q(k-1))) + V_F(k)
\end{align}
\vspace{-8pt}

Compared with SDC using conventional numerical integration rules, a notable feature of the Tustin transformation-based SDC is that \eqref{PI_Controller} does not involve the cumulative control error. For example, the trapezoidal rule-based  method~\cite{wang2020software} leads to a discretization of the PI controller as follows:
\vspace{-5pt}
\begin{align}
    g(n)&=K_pe_r(n)+T_sK_i\Big (\frac{e_r(0)+e_r(n)}{2} +\sum_{k=1}^{n-1}e_r(k)\Big )
    \label{Eq:PI_2}
\end{align}
For each time step, (\ref{Eq:PI_2}) needs to compute the cumulative control error ($\sum_{k=1}^{n-1}e_r(k)$), whereas (\ref{PI_Controller}) has no such requirements. 

As a result, the devised Tustin transformation-based SDC can ensure that the control performance is unaffected by historical errors and therefore provides improved numerical stability. In other words, the Tustin transformation-based SDC can eliminate the steady-state error even using moderate sampling rates, which significantly reduces the communication burden.

\subsubsection{Discretization of the Inertia Emulation Block}
This subsection derives the discretization rule of the outputs (i.e., phase angle $\delta$ and angular frequency $\omega$) of the inertia emulation block. 
Substituting~(\ref{Eq:Tustin}) into (\ref{Eq:rotor_trans}) yields the following:
\vspace{-5pt}
\begin{equation}
    \Theta(z) = \Big [ \frac{1}{2H(\frac{2}{T_s}\frac{z-1}{z+1})^2+D\frac{2}{T_s}\frac{z-1}{z+1}}\Big ] T(z)+\frac{1}{\frac{2}{T_s}\frac{z-1}{z+1}}W(z)
    \label{Eq:rotor_zform}
\end{equation}
where $\Theta(z)$ denotes the $z$ transform of $\delta$.
Normalize the constant term in the denominator of the transfer function in ~(\ref{Eq:rotor_zform}) to 1, and after rearrangement and let $K = 2/T_s$, we have:
\vspace{-5pt}
\begin{align} \label{Eq:rotor_zform2}
     \Theta(z)&=(1-\frac{4HK^2}{2HK^2+DK}z^{-1}+\frac{2HK^2-DK}{2HK^2+DK}z^{-2}) \notag \\ & +\frac{T(z)}{2HK^2+DK}(1+2z^{-1}+z^{-2}) +\frac{W(z)}{K}(1-z^{-2})
\end{align}
\vspace{-8pt}

Consequently, the discretized recursive equation of $\delta(n)$ is
\vspace{-5pt}
\begin{align}
     \delta(k) &=\frac{4HK^2}{2HK^2+DK}\delta(k-1)-\frac{2HK^2-DK}{2HK^2+DK}\delta(k-2) \notag \\ & +\frac{1}{2HK^2+DK} \Big [ T_e(k)+2T_e(k-1)+T_e(k-2) \Big ] \notag \\ & +\frac{1}{K}\Big [ \omega_s(k)-\omega_s(k-2) \Big ]
     \label{Eq:delta}
\end{align}

Then, $\omega$ can be readily obtained from the  derivative of $\delta$:
\begin{equation}
    \omega(k) = [\delta(k)-\delta(k-1)]/T_s
\end{equation}
\vspace{-10pt}

Other control blocks, such as the inner-loop controller, supplementary controllers, 
can also be discretized using the same procedure to conduct the software-based implementation. Due to page limitations, details are omitted here. Finally, \SDVSC is constructed by integrating all the discretized control blocks.

\vspace{-10pt}
\subsection{Programmable Management of \SDVSCs for Wind Farm Operations} \label{sec:management}

\newpage

Up to this point, we have achieved \SDVSC for an individual wind turbine. Further, in a wind farm, wind turbines equipped with \SDVSC can operate in parallel to increase the available \VSC capacity. This subsection discusses the management of multiple \SDVSCs to assist wind farm-level operations.

Because of the utilization of SDC, each wind turbine equipped with \SDVSC can flexibly switch its operating mode between the \VSC mode and various grid-following/grid-forming modes (e.g., $V/f$ control, $P/Q$ control, droop control). The wind farm control center (WFCC) manages the software-defined operations of wind turbines: 
\begin{itemize}[leftmargin=*]
\item Once a wind turbine is required to operate at the \VSC mode, the WFCC sends out control commands to the SDC unit to initialize the \SDVSC, which includes the operating mode switching command, the programmable parameters of the \VSC (e.g., emulated inertia and damping factor), and the supplementary control functions to be applied. 
\item Then, the SDC unit runs the discretized controllers (detailed in Subsection \ref{sec:SDVSC}) based on measurements received at each sampling time and outputs the control signals (e.g., signals for generating SPWM and adjusting pitch angle) to the destination wind turbine to perform the \VSC control. 
\item Additionally, the SDC unit can also generate backup controllers to provide control redundancy and better robustness. The backup controllers will immediately pop up once there is a failure in the master controllers. 
\end{itemize}

In summary, with \SDVSC, the SC functionalities become one type of built-in services that wind turbines can provide. \SDVSC enables two types of programmability that can not be achieved by traditional hardware-based SCs or \VSCs: 
i) all the control parameters can be programmed much more efficiently than hardware-inverter-based \VSC benefiting from the full softwarization; 
ii) the overall capability for providing the \VSC functionalities can also be flexibly programmed by switching the operating modes of wind turbines via SDC.
Such programmabilities, without restrictions of hardware dependence, make \SDVSC very adaptive to changing operating conditions and cost-effective for deployment and upgrading.

\vspace{-6pt}
\section{SDN-Enhanced \SDVSC Management for Large-Scale Wind Energy Integration}\label{sec:SDNCom}

Since \SDVSC management involves frequent communications between WFCC, wind turbines and the SDC unit, a dependable communication network is indispensable for guaranteeing the reliable operations of \SDVSCs. This section establishes a software-defined networking (SDN)-enabled communication scheme for \SDVSC management, which allows for ultra-reliable communication with low delays or congestion and hence supports resilient operations of large wind farms with hundreds or even thousands of wind turbines. 

\vspace{-6pt}
\subsection{Communication Activities in \SDVSC Management}

In the \SDVSC management procedure detailed in Subsection \ref{sec:management}, three major communication activities are involved:
\begin{itemize} [leftmargin=*]
\item \textbf{Monitoring}: For each wind turbine, its connection status (i.e., connected or exited) and the operation mode (i.e., \VSC mode or other modes) are continuously monitored and sent to the WFCC for event detection and decision making. 
\item \textbf{Measurements}: Once the \VSC mode is enabled for a specific wind turbine, measurements of this wind turbine, including the three-phase output voltages and currents, are sent to the SDC unit to implement the \SDVSC control.
\item \textbf{SDC/WFCC commands}: The SDC unit sends out the control signals (e.g., the modulation signals) to specific wind turbines. The WFCC may also send control commands, such as voltage references, to adjust the operation of \SDVSCs.
\end{itemize}

Traditional communication networks cannot fulfill \SDVSCs' data transmission requirements (i.e., low latency and high reliability) because of insufficient visibility to support real-time network monitoring and imprompt reactions to dynamic network conditions (e.g., congestion and link failure). 
SDN offers new insights into developing ultra-resilient, scalable and automatically configurable communication  by separating the control/data planes and exploiting network virtualization technologies~\cite{zhang2019enabling}.
It enables direct control and management of communication networks through various functions, such as dynamic routing and traffic prioritization, and allows deploying new applications at a faster rate~\cite{zhang2021networked}.

Inspired by the SDN philosophy, we perform \SDVSC communications through an SDN-based communication network. 

\vspace{-8pt}
\subsection{SDN-Enhanced \SDVSC Management}

\begin{figure}[!h]
  \centering
  \includegraphics[width=0.45\textwidth]{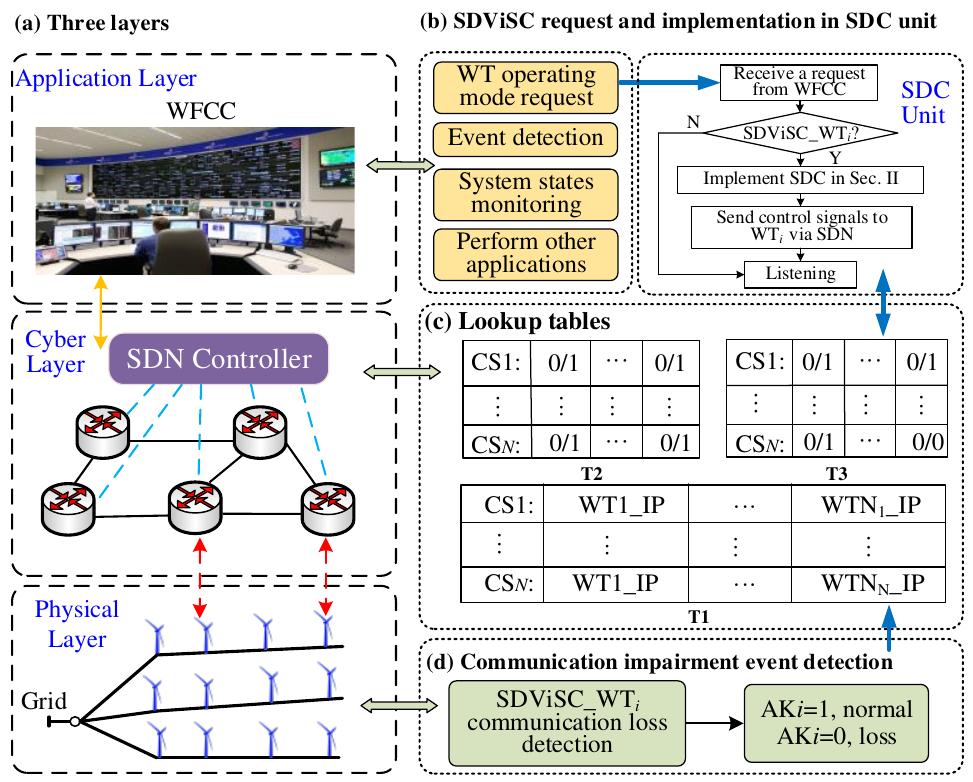}
  \vspace{-4pt}
    \caption{SDN-enhanced \SDVSC management.}
  \vspace{-8pt}
  \label{fig:Table_SDN} 
\end{figure}

The three-layered SDN-enhanced \SDVSC architecture is illustrated in Fig.~\ref{fig:Table_SDN}: 1) a physical layer containing wind turbines in wind farms, 2) a cyber layer where the SDN controller manages the communication network for \SDVSCs, and 3) an application layer for implementing \SDVSC as well as other grid-forming/grid-following controllers. 
\vspace{-5pt}

\subsubsection{Design of SDN Functions}

According to the communication activities involved in \SDVSC management, the SDN controller is designed to provide the following functions.

First, the SDN controller monitors the status of the cyber-physical system and dynamically updates the related information, including the connection status and operating modes of wind turbines, requests from WFCC, etc. Here, we design three lookup tables to store the monitored information (see details in Subsection \ref{sec:lookup table}): 1) the IP address table (T1), 2) the wind turbine's operating status and mode table (T2), and 3) the communication state table (T3). With those tables, the SDN controller can establish communication links as needed.

Second, the SDN controller manages the cyber system to ensure reliable and resilient communication to support \SDVSC operations. For example, it can build communication links between the SDC unit and wind turbines as needed according to the  information stored in T1, T2 and T3. It can also provide dynamic routing under network impairments, such as loss of communication. Here, we will design an event-based communication scheme to reduce communication burdens (see details in Subsection \ref{sec:event}).



\subsubsection{\SDVSC Network Monitoring}\label{sec:lookup table}

Typically, a large wind farm is divided into clusters, each consists of a group of wind turbines connected in series. We use the cluster (CS) number and the wind turbine number for indexing to design the lookup tables to store the communication network information. 

Using the IP protocol, each wind turbine has a unique IP address stored in T1 and retrieved by the SDN controller. T1 is updated when changes occur in the physical layer (e.g., plug-and-play of wind turbines) or in the communication configuration (e.g., adding, deleting, or modifying IP addresses). 

T2 stores wind turbines' operating modes using three two-digit numbers for each wind turbine: `0/0', `0/1', and `1/1'. The first digit indicates the current operating mode, where `1' represents the \VSC mode and `0' are other modes. The second digit indicates whether the \VSC function is enabled, where `0' means disabled (i.e., the wind turbine cannot work in the \VSC mode) and `1' means enabled (i.e., the wind turbine can choose to work in the \VSC mode). Definitely, only when the \VSC mode is enabled, the current operating mode can be 1.

T3 stores wind turbines' communication states also using three two-digit numbers: `0/0', `0/1', and `1/1'. The first digit represents the current communication state between a wind turbine and the SDC unit, where `1' means the communication is on and `0' means off. The second digit indicates whether there exists a link satisfying the communication requirement, where `1' means yes and `0' means no. 




\subsubsection{Event-Based Communication  for \SDVSC Management}\label{sec:event}


To reduce the communication bandwidth, we develop an event-based SDN communication scheme for \SDVSC management so that communication is only required under specific events.


The following discusses two typical events: 1) $\text{\SDVSC}\_\text{WT}_i$, where the SDN controller receives a request to operate the $i$-th wind turbine in the \SDVSC mode;  2) $\text{DR}_i$, where dynamic routing is required for the $i$-th wind turbine under communication impairments.


%
%


During wind farm operations, the SDN controller continuously listens to requests from WFCC. Once an $\text{\SDVSC}\_\text{WT}_i$ event is triggered, the SDN controller sends a command to the SDC unit and retrieves the IP address of the destination wind turbine from T1. Then, T2 and T3 are successively checked to confirm that the \VSC function is enabled and the corresponding communication state is on. Finally, the SDN controller creates flow tables for SDN switches to build communication links between the destination wind turbine and the SDC unit so that the \SDVSC control specified in Subsection \ref{sec:management} can be readily performed.

Under communication impairments, the dynamic routing event $\text{DR}_i$ will be triggered to switch to a new communication path to guarantee reliable data transmission. 
Although the SDN controller provides a built-in function to periodically detect the network status, its performance is limited by the detection period and thus may not respond timely. Therefore, to ensure fast dynamic routing, we define an acknowledge signal $\text{AK}_i$  to detect the network performance. If the $i$-th wind turbine  receives packets from the SDC unit normally, $\text{AK}_i$ is 1; otherwise, it changes to 0. Then, $\text{AK}_i$ is sent back to the SDN controller to check whether the dynamic routing should trigger by $\text{DR}_i = \text{\SDVSC}\_\text{WT}_i \land (^\neg \text{AK}_i)$. A delay with 0.04$s$ is added to the detection of $\text{DR}_i$ based on the delay effect analysis in~\cite{wang2020software}. The added delay ensures that false detection will not be induced because the communication delay of Ethernet switches is much smaller.   

In summary, the SDN-enhanced \SDVSC management offers three benefits unattainable by traditional communication techniques: 1) it can effectively manage the network configurations and perform dynamic routing under communication failures, which enables great flexibility, reliability and resilience for the communication between \SDVSCs, the WFCC and the SDC unit; 2) it adopts an event-based scheme, which avoids occupation of controller-to-switch bandwidth and reduces unnecessary data transmission; and 3) SDN benefits communications in complex networks by breaking communication barriers arising from proprietary protocols. This is because SDN directly controls the flow of data packets by creating the flow tables for routers among subnetworks, and adjusts data priority and throughput to avoid congestion. As such, \SDVSCs' deployment and management are not constrained by specific communication infrastructures and protocols. 

\vspace{-4pt}
\section{Experimental Results}
In this section, we thoroughly evaluate \SDVSC's performance  on a real-time and hardware-in-loop prototype using RTDS. First, we validate the Tustin transformation-based SDC algorithm. Second, the performance of the \SDVSC in comparison to the traditional SC is presented. Third, we validate the control efficacy of \SDVSC to support reliable and resilient operations of wind farms under various conditions. Last, we demonstrate the reliability and benefits of the SDN-enhanced communication scheme for \SDVSC management.

\vspace{-10pt}
\subsection{Testing Environment Setup}

\subsubsection{Test System}
\begin{figure}[!h]
  \centering
  \includegraphics[width=0.43\textwidth]{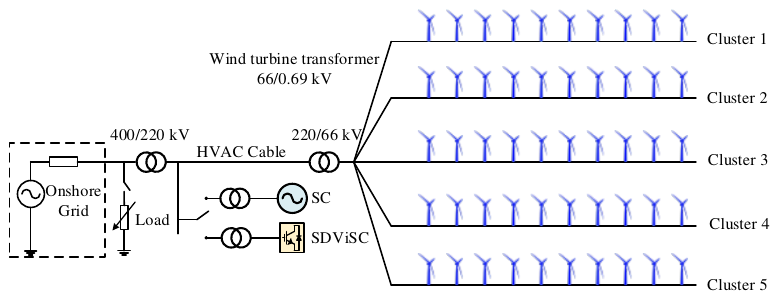}
  \vspace{-10pt}
  \caption{\label{fig:OWF} Topology of the test system: a 500MW OWF with 50 wind turbines.} 
\end{figure}

Fig.~\ref{fig:OWF} shows our designed 500 MW offshore wind farm (OWF) test system with 50 PMSG based wind turbines\footnote{The designed OWF is extended from the CIGRE C4.49 benchmark system, which has two equivalent wind turbines with aggregated models~\cite{kocewiak2020overview}.} to verify \SDVSC. 
The 50 wind turbines are grouped into 5 clusters and integrated into the onshore power grid through a high-voltage alternating current (HVAC) cable. For comparison studies, a 40 MVAR traditional SC (referred to~\cite{nguyen2018combination}) and an SDViSC with the same capacity by aggregating four wind turbines~\cite{kocewiak2020overview} are connected to the onshore point of coupling separately. Detailed designs are as follows:
\vspace{-4pt}
\begin{itemize}[leftmargin=*]
\item \textbf{Wind turbine controller}: Wind turbines originally adopt a grid-following-based double loop control. 
Detailed control topology and controller parameters are provided in~\cite{kocewiak2020overview}.
\item \textbf{Onshore grid}: It is modeled as an ideal voltage source connecting an impedance $R/X=0.1$ in series. The short-circuit ratio (SCR) can be adjusted to emulate different grid conditions~\cite{kocewiak2020overview} 
to provide a thorough evaluation of \SDVSC. 
\end{itemize}
\vspace{-8pt}
\vspace{-4pt}

\subsubsection{Real-Time Prototype}

As depicted in Fig.~\ref{fig:testbed}, the \SDVSC prototype has an OWF control center (OWFCC), an SDC unit, an SDN controller, four SDN switches, a traditional switch (L2 switch), GTNET$\times$2 cards, RTDS hardware, and the power system simulation software RSCAD that interacts with the RTDS hardware. Detailed settings are as follows:

\vspace{-10pt}
\begin{figure}[!h]
  \centering
  \includegraphics[width=0.43\textwidth]{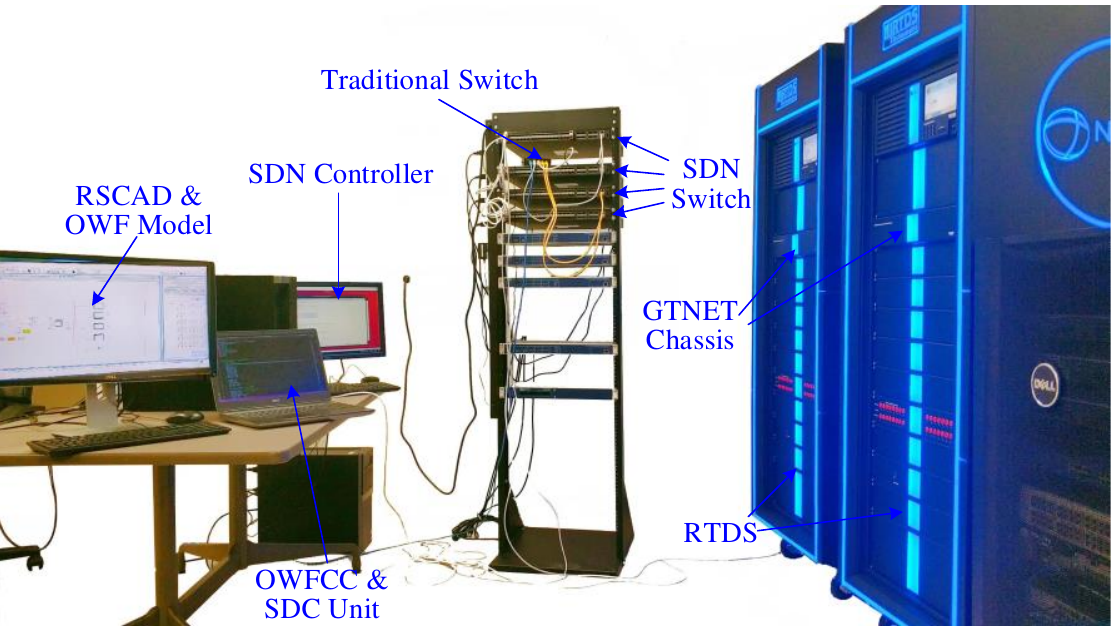}
  \vspace{-10pt}
  \caption{Real-time, cyber-in-the-loop prototype for \SDVSC.}
  \vspace{-6pt}
  \label{fig:testbed}
\end{figure}

\begin{itemize}[leftmargin=*]
\item \textbf{Real-time OWF simulation}: The OWF system (see Fig.~\ref{fig:OWF}) is developed and compiled in RSCAD. Two RTDS racks with 10 cores are utilized to simulate the OWF in real time. 
\item \textbf{Communication setup}: GTNET$\times$2 cards are utilized for RTDS hardware to communicate with external devices, such as the OWFCC and the SDC unit, over a LAN/WAN. 
The GTNET$\times$2 cards provide Gigabit Ethernet ports for IP-based communications. UDP is used to transmit packets.
\item \textbf{SDN setup}:  A real SDN network with an SDN controller and four SDN switches running OpenFlow protocols are used to manage the communication network. 
\end{itemize}
\vspace{-8pt}

\subsection{Numerical Stability of the Devised SDC Algorithm}

This part verifies the numerical stability of our designed Tustin transformation-based SDC algorithm and demonstrates its superiority over the trapezoidal rule-based SDC algorithm~\cite{wang2020software}. \VSCs directly implemented in RSCAD emulate the performance of hardware-dependent controllers and serve as the ground truth for evaluating the performance of software-defined controllers. A phase-to-ground fault is studied, which occurs at the grid side at 1s and is cleared at 1.3s.

\subsubsection{Performance of Hardware-Dependent \VSC}

\begin{figure}[!h]
  \centering
  \includegraphics[width=0.44\textwidth]{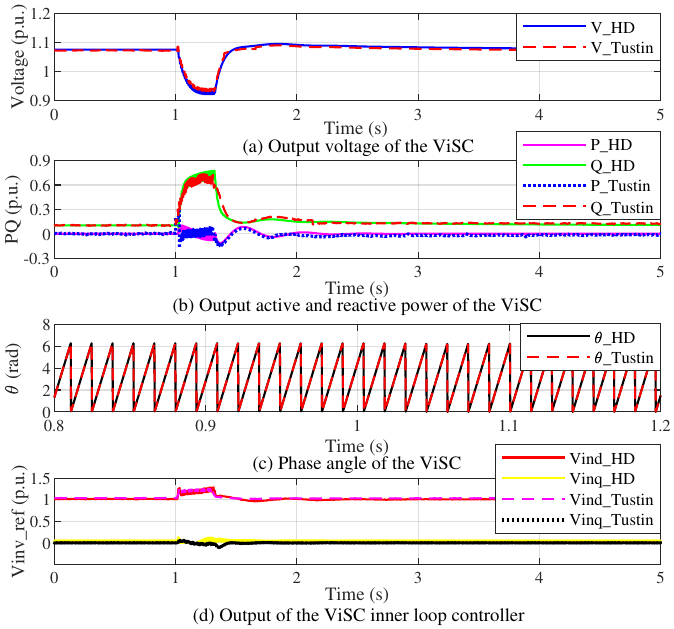}
  \vspace{-10pt}
  \caption{Performance of \VSC using Tustin transformation-based SDC (sampling time: 0.67ms) and its comparison with hardware-dependent (HD) \VSC.}
  \vspace{-4pt}
  \label{fig:study1_voltage}
\end{figure}

The solid lines in Fig.~\ref{fig:study1_voltage} illustrate the responses of hardware-dependent \VSCs. Fig.~\ref{fig:study1_voltage} (a) and (b) show that when the fault occurs, \VSCs immediately generate reactive power to support the grid voltage and successfully bring the system back to normal operation after the fault clearance. The phase angle $\theta$ in Fig.~\ref{fig:study1_voltage}(c) from the inertial emulation block and the voltage reference in Fig.~\ref{fig:study1_voltage}(d) for inner-loop modulation also demonstrate \VSCs function well as designed, i.e., providing inertia response and reactive power support as SCs do. 
\vspace{-8pt}

\subsubsection{Performance of \SDVSC based on Tustin Transformation}

The dash lines in Fig.~\ref{fig:study1_voltage} illustrate the performance of \SDVSC using the Tustin transformation-based SDC algorithm. A moderate sampling rate, i.e., 0.67ms, is adopted. Simulation shows \SDVSC performs as well as the hardware-dependent \VSC, and there is almost no obvious performance deterioration. Therefore, our method can facilitate fast and low-cost deployment of SC functionalities with qualified control performance. 




%
%

\begin{figure}[!h]
  \centering
  \includegraphics[width=0.44\textwidth]{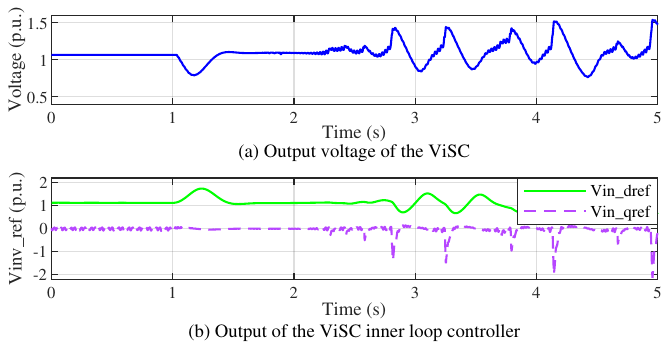}
  \vspace{-10pt}
  \caption{Performance 
  trapezoidal rule-based SDC (sampling time: 0.25 ms).}
  \label{fig:study1_VSC_4000}
\end{figure}

In contrast, Fig.~\ref{fig:study1_VSC_4000} presents the performance of the trapezoidal rule-based SDC algorithm~\cite{wang2020software} under a sampling rate 0.25 ms (i.e., the smallest sampling rate that can be realized under the communication bandwidth setting). It can be seen that the system collapses at around 2.2s, which indicates the trapezoidal-based SDC fails to provide the designed control functionality under limited communication bandwidth because of the accumulated historical error (as discussed in Subsection \ref{sec:Discretization}).
Although further decreasing the sampling rate can improve the performance of trapezoidal-based SDC, it will unavoidably bring higher requirements on communication and increased vulnerability against network impairments.
\vspace{-8pt}
\vspace{-2pt}

As a conclusion, the Tustin transformation-based SDC ensures enhanced numerical stability compared over existing methods, evidenced by the fact that a much larger sampling rate (lower communication requirement) can be adopted without affecting the control performance.
\vspace{-18pt}

\subsection{Comparison between the \SDVSC and Traditional SC}

This subsection compares their performance from four aspects: voltage regulation capability, inertia support, short-circuit capacity contribution and grid strength enhancement.
\subsubsection{Voltage Regulation Capability}
\begin{figure}[!h]
  \centering
  \includegraphics[width=0.434\textwidth]{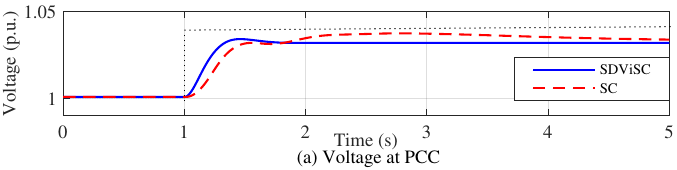}
  \includegraphics[width=0.434\textwidth]{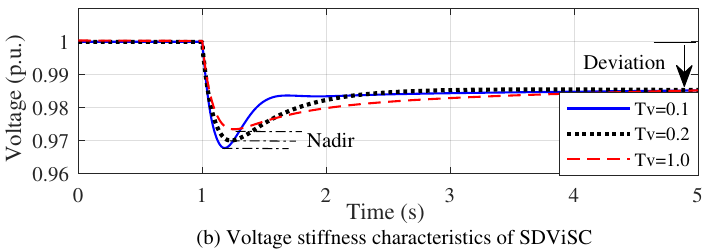}
  \vspace{-10pt}
  \caption{Voltage regulation performance and stiffness of \SDVSC under a strong onshore grid (SCR=7.14, AVR droop coefficients are both 0.02)}.
  \label{fig:Compare_Volg}
\end{figure}

Fig.~\ref{fig:Compare_Volg} (a) presents the voltage regulation performance when changing the voltage references for the AVRs of the traditional SC and \SDVSC from 1.0 p.u. to 1.04 p.u. at 1s separately. It can be seen that: 1) both the SC and \SDVSC boost the voltage around 1.035 p.u. finally; 2) \SDVSC offers faster regulation speed and thus improved voltage regulation capability than the traditional SC.

Fig.~\ref{fig:Compare_Volg} (b) shows the \SDVSC's voltage stiffness characteristics, the inner potential’s ability to tolerate the difference between the real and reference reactive power over a time interval. It is determined by $T_v=1/K_{iv}$, the reciprocal of the integrator parameter of AVR's PI controller~\cite{shang2022voltage}. The grid voltage decreases by 0.08 p.u at 1 s, it can be observed that: 1) with a larger $T_v$, the changing rate of voltage at the onshore point of coupling is reduced; 2) the static deviation of voltage is not improved; and 3) the voltage nadir is pulled up and enhanced. This kind of stiffness characteristic provided by the \SDVSC is beneficial to improve the voltage dynamics.
\vspace{-6pt}
\vspace{-6pt}

\subsubsection{Inertia Support}
\begin{figure}[!h]
  \centering
  \includegraphics[width=0.434\textwidth]{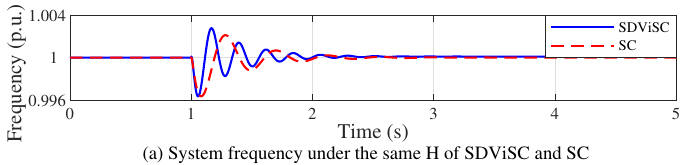}
  \includegraphics[width=0.434\textwidth]{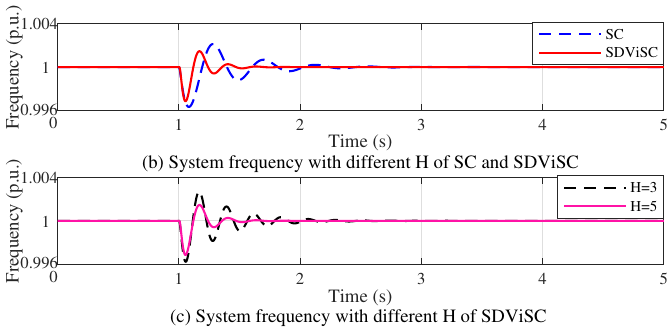}
  \vspace{-8pt}
  \caption{System responses with \SDVSC and SC under a strong grid (SCR=7.14). (a) with the same H = 3s, (b) with H =3s of SC and H = 5s of SDViSC; (c) Programmable H=3s and H =5s for SDViSC.}
  \label{fig:Compare_inertia1}
\end{figure}

\begin{figure}[!h]
  \centering
  \includegraphics[width=0.434\textwidth]{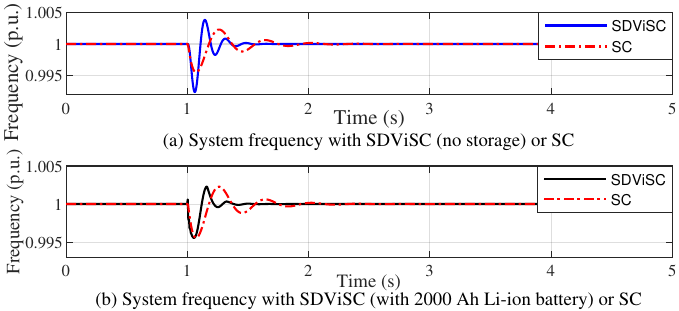}
  \vspace{-8pt}
  \caption{System frequency using \SDVSC  (without and with storage) and SC with same  inertia constant $H = 4$ under a strong grid (SCR=7.14).}
  \label{fig:Compare_inertia3}
\end{figure}

As shown in Fig.~\ref{fig:MSC_Control}, the machine side converter is controlled to inject current to the DC link capacitor to maintain a constant voltage during the inertia support process. Therefore, the energy stored in the capacitor (determined by capacitance value) and extracted from the machine side impacts the performance of \SDVSC. Here two load-increasing cases are performed:

\begin{itemize} [leftmargin=*]
    \item i) Load increases by 50 MW: Fig.~\ref{fig:Compare_inertia1} (a) illustrates the inertia support performance when the \SDVSC and traditional SC have the same inertia constant (i.e., H=3s). The system frequency decreases as the load increases and the frequency nadir is almost the same for the \SDVSC and SC. 
    Fig.~\ref{fig:Compare_inertia1} (b) and (c) further present the advantage of the programmability of \SDVSC in comparison to SCs. The frequency nadir is decreased when increasing the inertia of \SDVSC, which clearly shows its advantage to adapt to the evolution of the grid's needs. The decreased oscillation frequency also indicates enhanced system performance.
    \item ii) Load increases by 100 MW: Fig.~\ref{fig:Compare_inertia3} (a) demonstrates the system responses with the \SDVSC and traditional SC of the same capacity and inertia constant. It can be seen that the system experiences a larger frequency dip with the SDViSC than that with SC because of the limited stored energy by wind turbines. Fig.~\ref{fig:Compare_inertia3} (b) presents the frequency responses when paralleling a 2000 Ah Li-ion battery with the DC capacitor. The obviously decreased frequency nadir indicates improved inertia support in comparison with the case having no storage.
\end{itemize}
\vspace{-5pt}
\vspace{-8pt} 
\vspace{-6pt}

As a conclusion, equivalent inertia support as traditional SCs can be offered by the \SDVSC when having adequate stored energy~\cite{ding2021two}. It is also worth highlighting that the inertia constant of \SDVSC can be programmed to satisfy the system's evolving needs. Guaranteeing a solid inertia support service similar to traditional SCs, additional storage equipment is required to be installed or operating multiple WTs as \SDVSCs in parallel, which will be validated in subsection \ref{sec:parallel_SDViSC}.


\subsubsection{Short-Circuit Capacity Contribution}

\begin{figure}[!h]
  \centering
  \includegraphics[width=0.43\textwidth]{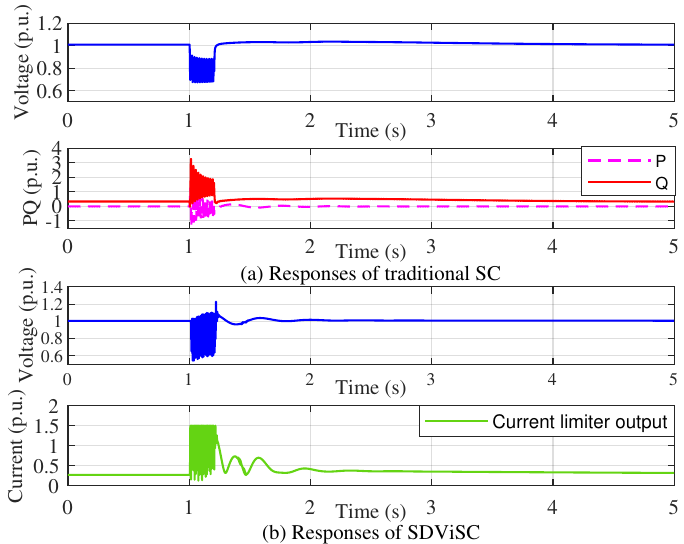}
  \vspace{-8pt}
  \caption{System responses under grid side faults (SCR=7.14).}
  \label{fig:Compare_ShortC}
\end{figure}

The maximum peak current is set to be 1.5 p.u.~\cite{lepour2021performance}.  Fig.~\ref{fig:Compare_ShortC} presents the outputs of the \SDVSC and the traditional SC under a single-phase to ground fault, which starts at 1 s and is cleared at 1.2 s. It can be observed from the results that:
\begin{itemize} [leftmargin=*]
    \item Both the \SDVSC and the traditional SC instantaneously inject reactive power to boost the voltage. 
    \item As shown in  Fig.~\ref{fig:Compare_ShortC} (a), the SC offers a short-circuit current that rises to 4 p.u. and the residual voltage is above 0.7 p.u.. 
    \item The \SDVSC's contribution is directly limited by its overcurrent capability, as shown in Fig.~\ref{fig:Compare_ShortC} (b). The residual voltage is about 0.6 p.u., which is lower than that of the SC. Therefore, an \SDVSC with a larger capacity is required to provide comparable short-circuit current as the SC does. 
    \vspace{-6pt}
\end{itemize}
\vspace{-10pt}
\vspace{-6pt}

\subsubsection{Grid Strength Enhancement}

\begin{figure}[!h]
  \centering
  \includegraphics[width=0.44\textwidth]{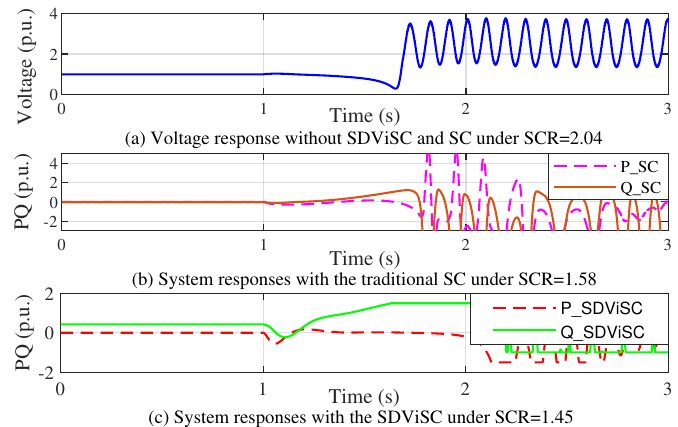}
  \vspace{-8pt}
  \caption{System responses under weak grids emulated by changing the equivalent impedance (or SCR) of the onshore grid at 1 s.}
  \label{fig:Compare_Weak}
\end{figure}

We compare the capability to suppress instability under weak grid conditions via three cases: 1) without \SDVSC or SC, 2) with only SC and 3) with only \SDVSC. Fig.~\ref{fig:Compare_Weak} presents the results that the studied case is becoming unstable  under the largest SCR (the boundary value for system stability and instability). It can be seen that:
\begin{itemize} [leftmargin=*]
    \item When there is no SC or \SDVSC, the system is unstable (see Fig.~\ref{fig:Compare_Weak} (a)) when the SCR decreases to 2.04, the voltage oscillates severely and deviates largely from the rated value. 
    \item With SC, the system is stable (see Fig.~\ref{fig:Compare_Weak} (b)) under weaker grid conditions (lower SCR) compared with Fig.~\ref{fig:Compare_Weak} (a), which shows that the weak grid is indeed enhanced. The system still loses stability when the SCR decreased to 1.58 due to the limited capability of SC.
    \item With \SDVSC, the system can be stabilized under further lower SCR (see Fig.~\ref{fig:Compare_Weak} (c)). Thus, the grid-forming control based \SDVSC with changeable parameters performs better than SCs to maintain stability under changing conditions.
\end{itemize}
\vspace{-12pt}
\vspace{-8pt}

In summary, the following observations can be obtained:
\begin{itemize} [leftmargin=*]
    \item The \SDVSC outperforms the SC in terms of voltage regulation speed and its voltage stiffness characteristics are beneficial to system voltage dynamics. 
    \item With adequate stored energy, \SDVSC can provide equivalent inertia support as the SC, alternative ways are installing extra storage equipment or operating multiple WTs as \SDVSCs in parallel to reduce the effects of hardware limitation. Its parameters can be flexibly reconfigured to adapt to changing grid conditions.  
    \item The overcurrent capability of converters, which is definitely below the capability of SC, limits their short-circuit capacity contribution. Quantification for specific converters may be needed before deployment to provide the same contribution. 
    \item The \SDVSC with programmable parameters performs better given enough capacity than the SC to maintain transient stability under weak grid conditions.
\end{itemize}

\vspace{-12pt}
\subsection{Control Efficacy of \SDVSC}\label{sec:parallel_SDViSC}

We verify \SDVSC's control efficacy under three scenarios, including plug-and-play of \SDVSCs, parallel operation of \SDVSCs and oscillation damping under weak grid conditions. 

\subsubsection{Plug-and-Play of \SDVSCs} 

In this part, we demonstrate how the OWF reacts under the plug-and-play operation of \SDVSCs. 
Specifically, two wind turbines in Cluster 1 and Cluster 4 are plugged into the OWF as \SDVSCs at 1.2 s and 4.8 s, respectively. 
Fig.~\ref{fig:study2_Plug} presents the physical and cyber states of these two \SDVSCs before and after the plug operation. 
It can be observed from the results that: 
\vspace{-8pt}
\begin{figure}[!h]
  \centering
  \includegraphics[width=0.44\textwidth]{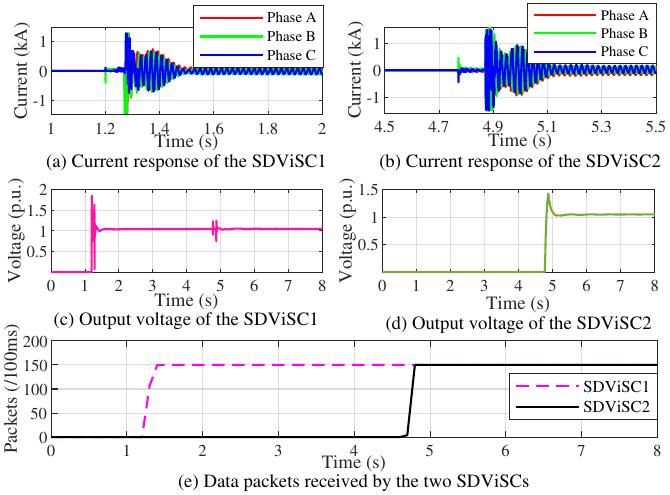}
  \vspace{-8pt}
  \caption{Physical and cyber states of \SDVSCs under plug-and-play.}
  \vspace{-8pt}
  \label{fig:study2_Plug}
\end{figure}

  

\begin{itemize} [leftmargin=*]
    \item When \SDVSCs are connected to the OWF, the software-defined controllers immediately start to operate the wind turbines as \VSCs to provide voltage and inertia support. 
    \item As shown in  Fig.~\ref{fig:study2_Plug}(a)-(d), currents and voltages of \SDVSCs can be stabilized in a short time (i.e., within around 0.1s) after the plug operation, which indicates a satisfactory transient performance of the designed \VSC control. 
    \item Communication is only required when the wind turbine is operating as a \VSC. Fig.~\ref{fig:study2_Plug} (e) presents the packets received by the two \SDVSCs, which clearly show that the SDN controller builds the communication link in a timely manner after receiving a request from the OWFCC. 
\end{itemize}

\subsubsection{Parallel Operation of \SDVSCs for Voltage Regulation}


A single \SDVSC may not be able to provide the functionalities of a large-capacity SC.
Therefore, we propose to operate multiple \SDVSCs in parallel to increase the overall capacity. 

First, we demonstrate the efficacy of the parallel operation of \SDVSCs for providing voltage regulation services. The leftmost 4 wind turbines in Cluster 3 operate in parallel as \SDVSCs with voltage droop coefficients of 0.03, 0.05, 0.07, and 0.1. Results are shown in Fig.~\ref{fig:study2_VSC_reg}:
\vspace{-8pt}
\begin{figure}[!h]
  \centering
  \includegraphics[width=0.44\textwidth]{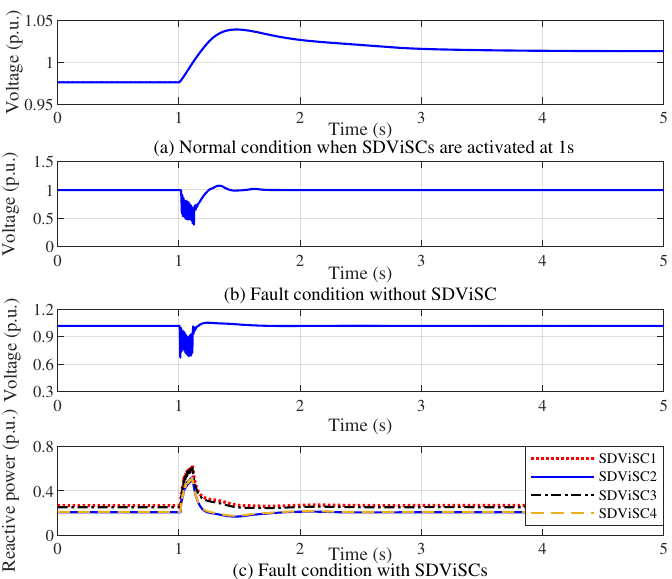}
  \vspace{-8pt}
  \caption{System responses when in-parallel \SDVSCs are activated at 1s.}
  \vspace{-8pt}
  \label{fig:study2_VSC_reg}
\end{figure}
\begin{itemize}[leftmargin=*]
\item Fig.~\ref{fig:study2_VSC_reg} (a) first studies the performance of \SDVSCs in normal operating conditions (i.e., without any faults). It can be seen that before the deployment of \SDVSC (before 1s), the PCC voltage is 0.97 p.u., which is below the rated voltage. After \SDVSCs are activated to inject reactive power, the PCC voltage is accordingly boosted to 1.02 p.u..
\item Fig.~\ref{fig:study2_VSC_reg} (b)-(c) further study the system responses under a phase-to-ground fault that occurred at the grid side at 1s.
Fig.~\ref{fig:study2_VSC_reg} (b) shows that without \SDVSCs, the PCC voltage drops to 0.4 p.u. after the fault. However, as shown in Fig.~\ref{fig:study2_VSC_reg} (c), when \SDVSCs are deployed, the PCC voltage during and after the fault can be significantly improved. The reason is that \SDVSCs perform reactive power sharing according to  their voltage droop coefficients to support the grid voltages. 
\end{itemize}
\vspace{-4pt}

In summary, simulation results indicate that \SDVSCs are competent to provide voltage regulation functionality as traditional SCs under both normal operating and fault conditions.

%
%


\subsubsection{Parallel Operation of \SDVSCs for Improving the Transient Stability of Weak Grids} 

Further, we demonstrate the capability of \SDVSCs to improve the system's transient stability under weak grid conditions. At time $t=1s$, the SCR is changed from 7.14 to 1.51 by adjusting the impedance of the onshore grid to simulate weak grid conditions. 
\begin{figure}[!t]
  \centering
  \includegraphics[width=0.44\textwidth]{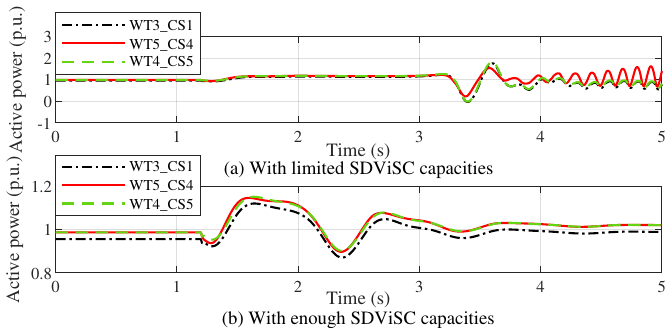}
  \vspace{-8pt}
  \caption{System response during the transition between strong and weak grid conditions. 
  WT1\_CS1: Wind Turbine 1 in Cluster 1.}
  \vspace{-8pt}
  \label{fig:study2_Weak_noVSC}
\end{figure}

In comparison to the scenario without \SDVSCs shown in Fig.~\ref{fig:Compare_Weak} (a), Fig.~\ref{fig:study2_Weak_noVSC} compares the system performance with 3 \SDVSCs (i.e., limited \VSC capacities) and 4 \SDVSCs (i.e., enough \VSC capacities). It can be observed that:
\vspace{-4pt}
\begin{itemize} [leftmargin=*]
\item With 3 wind turbines operating as \SDVSCs, Fig.~\ref{fig:study2_Weak_noVSC} (a) shows that \SDVSCs indeed enhance the system's transient stability compared with Fig.~\ref{fig:Compare_Weak} (a), and hence this weak grid can maintain stable until 3s. However, because of the limited capacity of \SDVSCs, the system still deviates from the nominal operating point after 3s and finally loses stability.
\item Further, with 4 \SDVSCs, Fig.~\ref{fig:study2_Weak_noVSC} (b) shows that the oscillations can be damped and the system can finally achieve a stable operating condition. 
\end{itemize}
\vspace{-4pt}

In summary, the ability of \SDVSCs to improve transient stability can be flexibly programmed by switching the operating modes of wind turbines and changing the number of available \VSC controllers. Therefore, \SDVSCs can adaptably satisfy the system needs under changing grid conditions, which outperforms traditional SCs with nonadjustable capacity.

\vspace{-8pt}
\subsection{Communication Resilience of \SDVSC}
This subsection demonstrates the communication resilience of the designed SDN-enhanced \SDVSC management. Two typical cyber events are studied: 1) communication network impairment and 2) controller failover.

\subsubsection{Network Impairment} 

\begin{figure}[!t]
  \centering
  \includegraphics[width=0.44\textwidth]{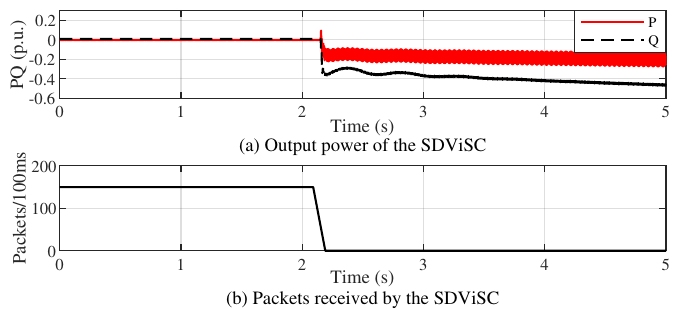}
  \vspace{-8pt}
  \caption{System response without SDN-enabled dynamic routing.}
  \label{fig:study3_NO_DR}
\end{figure}
\begin{figure}[!t]
  \centering
  \includegraphics[width=0.44\textwidth]{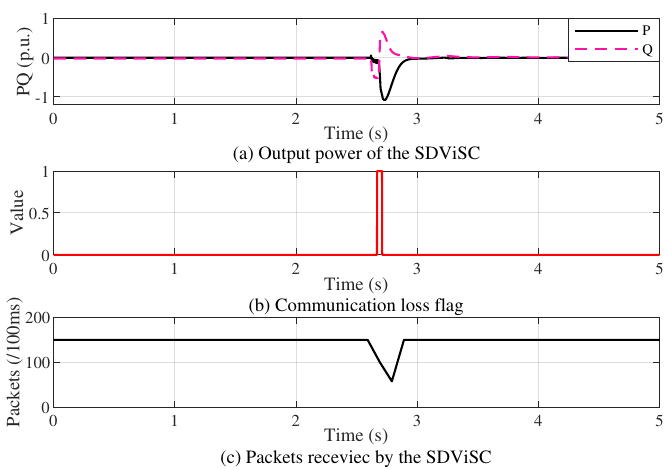}
  \vspace{-8pt}
  \caption{System response with SDN-enabled dynamic routing.}
  \vspace{-8pt}
  \label{fig:study3_With_DR}
\end{figure}

First, we study the performance of \SDVSC under network impairments, which may occur frequently in real-world communication networks and can significantly affect the operation of \SDVSCs. A physical failure of the communication link of wind turbine 1 in Cluster 1 is triggered to simulate bad network conditions. The testing results without and with the SDN-enabled communication are shown in Figs.~\ref{fig:study3_NO_DR} and \ref{fig:study3_With_DR}, respectively. It can be observed that:
\begin{itemize}[leftmargin=*]
    \item Without the SDN-based communication scheme, the communication is interrupted immediately when a link fails (see Fig.~\ref{fig:study3_NO_DR} (b)), and the control signals cannot be transmitted to the \SDVSC. Consequently, the \SDVSC becomes abnormal without reliable communication (see Fig.~\ref{fig:study3_NO_DR} (a)).
    \item With the SDN-based communication scheme, the SDN controller detects a communication loss event (see Fig.~\ref{fig:study3_With_DR} (b)), and dynamic routing is performed to switch to another route. Fig.~\ref{fig:study3_With_DR} (c) shows that the data transmission quickly recovers. The \SDVSC becomes normal after a short transient process.
\end{itemize}





\begin{figure}[!t]
  \centering
  \includegraphics[width=0.45\textwidth]{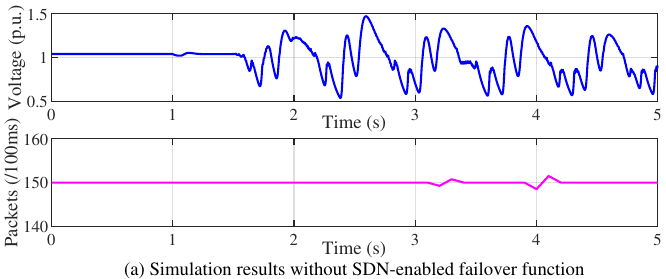}
  \includegraphics[width=0.45\textwidth]{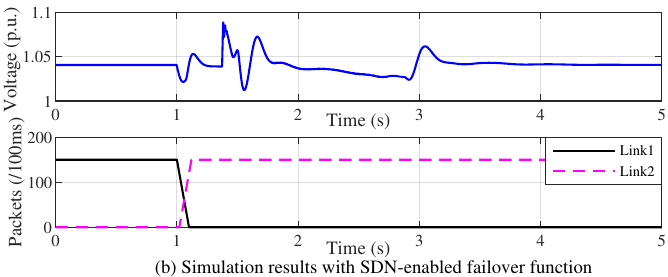}
  \vspace{-8pt}
  \caption{\SDVSC's response with and without SDN-enabled failover function.}
  \vspace{-8pt}
  \label{fig:study3_NO_Failover}
\end{figure}
\vspace{-4pt}

\subsubsection{Controller Failover} 
One salient feature of the developed SDN-enhanced \SDVSC is the controller failover function, which ensures during any controller failure, a backup controller can be conveniently implemented using software to provide redundancy and reliability for wind farm operations. 

Fig.~\ref{fig:study3_NO_Failover} provides the results with and without the SDN-enabled controller failover function carried out for wind turbine 1 in Cluster 1. It can be seen that:
\begin{itemize} [leftmargin=*]
    \item When the SDN-enabled failover function is disabled, the performance of the \SDVSC deteriorates severely after the controller malfunction, even though the cyber network is still working (see Fig.~\ref{fig:study3_NO_Failover} (a)). 
    \item When the failover function is enabled, the backup controller immediately starts to work after the controller malfunction, and the SDN controller establishes a new communication link (see Fig.~\ref{fig:study3_NO_Failover} (b)). The operation of \SDVSC becomes normal after a short period of oscillations. The handover capability enabled by the SDN-based communication scheme ensures the stable functioning of \SDVSC.
\end{itemize}

\vspace{-6pt}

\section{Conclusion}
In this paper, we devise \SDVSC to operate full converter wind turbines as virtual SCs in an unprecedented programmable, adaptable and lightweight manner. A Tustin transformation-based SDC algorithm is developed to virtualize controllers with fast dynamics in \SDVSC. An SDN-enabled communication scheme is further established for \SDVSC management with great communication resilience and reduced communication burden. A real-time, cyber-in-the-loop prototype for a large wind farm is built to validate the efficacy of \SDVSC in an RTDS environment. Extensive experimental results demonstrate that \SDVSC can be operated flexibly to provide SC capabilities to support reliable and resilient offshore wind farm operations. Comparison results with SCs show that the hardware limitation can be reduced by using together with storage and enhanced power converters.

\vspace{-10pt}

\bibliographystyle{IEEEtran}
\bibliography{mybibfile}

\begin{thebibliography}{10}
\providecommand{\url}[1]{#1}
\csname url@samestyle\endcsname
\providecommand{\newblock}{\relax}
\providecommand{\bibinfo}[2]{#2}
\providecommand{\BIBentrySTDinterwordspacing}{\spaceskip=0pt\relax}
\providecommand{\BIBentryALTinterwordstretchfactor}{4}
\providecommand{\BIBentryALTinterwordspacing}{\spaceskip=\fontdimen2\font plus
\BIBentryALTinterwordstretchfactor\fontdimen3\font minus \fontdimen4\font\relax}
\providecommand{\BIBforeignlanguage}[2]{{%
\expandafter\ifx\csname l@#1\endcsname\relax
\typeout{** WARNING: IEEEtran.bst: No hyphenation pattern has been}%
\typeout{** loaded for the language `#1'. Using the pattern for}%
\typeout{** the default language instead.}%
\else
\language=\csname l@#1\endcsname
\fi
#2}}
\providecommand{\BIBdecl}{\relax}
\BIBdecl

\bibitem{sunrise}
``{Offshore Wind Power for New York},'' \url{https://sunrisewindny.com/}, 2022.

\bibitem{liu2022open}
M.~V. Liu and \emph{et al}, ``An open source representation for the {NYS} electric grid to support power grid and market transition studies,'' \emph{IEEE Trans. Power Syst.}, 2022.

\bibitem{wan2022cooperative}
W.~Wan, P.~Zhang, M.~A. Bragin, and P.~B. Luh, ``Cooperative fault management for resilient integration of renewable energy,'' \emph{Electr. Power Syst. Res.}, vol. 211, p. 108147, 2022.

\bibitem{wu2019impedance}
H.~Wu, X.~Wang, and {\L}.~H. Kocewiak, ``Impedance-based stability analysis of voltage-controlled mmcs feeding linear ac systems,'' \emph{IEEE J. Emerg. Sel. Top. Power Electron.}, vol.~8, no.~4, pp. 4060--4074, 2019.

\bibitem{bakhshizadeh2018improving}
M.~K. Bakhshizadeh and \emph{et al}, ``Improving the impedance-based stability criterion by using the vector fitting method,'' \emph{IEEE Trans. Energy Convers.}, vol.~33, no.~4, pp. 1739--1747, 2018.

\bibitem{li2022dominant}
C.~Li, S.~Wang, F.~Colas, and J.~Liang, ``Dominant instability mechanism of vsi connecting to a very weak grid,'' \emph{IEEE Trans. Power Syst.}, vol.~37, no.~1, pp. 828--831, 2022.

\bibitem{nguyen2018combination}
H.~T. Nguyen, G.~Yang, A.~H. Nielsen, and P.~H. Jensen, ``Combination of synchronous condenser and synthetic inertia for frequency stability enhancement in low-inertia systems,'' \emph{IEEE Trans. Sustain. Energy.}, vol.~10, no.~3, pp. 997--1005, 2019.

\bibitem{nguyen2020applying}
H.~T. Nguyen and \emph{et al}, ``Applying synchronous condenser for damping provision in converter-dominated power system,'' \emph{J. Modern Power Syst. Clean Ener.}, vol.~9, no.~3, pp. 639--647, 2020.

\bibitem{jia2018impact}
J.~Jia and \emph{et al}, ``Impact of {VSC} control strategies and incorporation of synchronous condensers on distance protection under unbalanced faults,'' \emph{IEEE Trans. Ind. Electron.}, vol.~66, no.~2, pp. 1108--1118, 2018.

\bibitem{wang2020impact}
Y.~Wang, L.~Wang, and Q.~Jiang, ``Impact of synchronous condenser on sub/super-synchronous oscillations in wind farms,'' \emph{IEEE Trans. Power Deliv.}, vol.~36, no.~4, pp. 2075--2084, 2020.

\bibitem{nguyen2021technical}
H.~T. Nguyen, M.~N. Chleirigh, and G.~Yang, ``A technical \& economic evaluation of inertial response from wind generators and synchronous condensers,'' \emph{IEEE Access}, vol.~9, pp. 7183--7192, 2021.

\bibitem{bao2022wind}
L.~Bao and \emph{et al}, ``Wind farms in weak grids stability enhancement: Syn{C}on or {STATCOM}?'' \emph{Electr. Power Syst. Res.}, vol. 202, 2022.

\bibitem{kocewiak2020overview}
L.~Kocewiak and \emph{et al}, ``Overview status and outline of stability analysis in converter-based power systems,'' in \emph{Proc. Virtual Wind Integr. Workshop}, 2020, p.~10.

\bibitem{wang2020software}
L.~Wang, Y.~Qin, Z.~Tang, and P.~Zhang, ``Software-defined microgrid control: The genesis of decoupled cyber-physical microgrids,'' \emph{IEEE Open Access J. power energy}, vol.~7, pp. 173--182, 2020.

\bibitem{mo2017evaluation}
O.~Mo, S.~D'Arco, and J.~A. Suul, ``Evaluation of virtual synchronous machines with dynamic or quasi-stationary machine models,'' \emph{IEEE Trans. Industr. Electron.}, vol.~64, no.~7, pp. 5952--5962, 2017.

\bibitem{taul2019current}
M.~G. Taul and \emph{et al}, ``Current limiting control with enhanced dynamics of grid-forming converters during fault conditions,'' \emph{IEEE J. Emerg. Sel. Top. Power Electron.}, vol.~8, no.~2, pp. 1062--1073, 2019.

\bibitem{chinchilla2006control}
M.~Chinchilla and \emph{et al}, ``Control of permanent-magnet generators applied to variable-speed wind-energy systems connected to the grid,'' \emph{IEEE Trans. Energy Convers.}, vol.~21, no.~1, pp. 130--135, 2006.

\bibitem{lyu2019coordinated}
X.~Lyu, J.~Zhao, Y.~Jia, Z.~Xu, and K.~P. Wong, ``Coordinated control strategies of {PMSG}-based wind turbine for smoothing power fluctuations,'' \emph{IEEE Trans. Power Syst.}, vol.~34, no.~1, pp. 391--401, 2019.

\bibitem{zhao2021novel}
R.~Zhao and \emph{et al}, ``A novel discretization method for multiple second-order generalized integrators,'' \emph{IEEE Trans. Power Electron.}, vol.~36, no.~10, pp. 10\,998--11\,002, 2021.

\bibitem{al2007novel}
M.~Al-Alaoui, ``Novel approach to analog-to-digital transforms,'' \emph{IEEE Trans. Circuits Syst. I: Regul. Pap.}, vol.~54, no.~2, pp. 338--350, 2007.

\bibitem{zhang2019enabling}
P.~Zhang, B.~Wang, P.~B. Luh, L.~Ren, and Y.~Qin, ``Enabling resilient microgrid through ultra-fast programmable network,'' Dec.~10, 2019, {US Patent 10,505,853}.

\bibitem{zhang2021networked}
P.~Zhang, \emph{Networked microgrids}.\hskip 1em plus 0.5em minus 0.4em\relax Cambridge University Press, 2021.

\bibitem{shang2022voltage}
L.~Shang and \emph{et al}, ``{VSC}--based voltage stiffness compensator to improve grid voltage dynamics,'' \emph{Front. Energy Res.}, p.~3, 2022.

\bibitem{ding2021two}
T.~Ding and \emph{et al}, ``Two-stage chance-constrained stochastic thermal unit commitment for optimal provision of virtual inertia in wind-storage systems,'' \emph{IEEE Trans. Power Syst.}, vol.~36, no.~4, pp. 3520--3530, 2021.

\bibitem{lepour2021performance}
D.~Lepour and \emph{et al}, ``Performance assessment of synchronous condensers vs voltage source converters providing grid-forming functions,'' in \emph{2021 IEEE Madrid PowerTech}.\hskip 1em plus 0.5em minus 0.4em\relax IEEE, 2021, pp. 1--6.

\end{thebibliography}



\end{document}